# Impact of Business technologies on the success of E-commerce Strategies: SMEs Perspective

**Ziad Hmwd A Almtiri**
*VU Business School,* Victoria University,
Melbourne, Australia,
Email:
Ziad_Amtiri@outlook.com

**Shah Jahan Miah**
*Newcastle Business school*
The University of Newcastle NSW, Australia
Email: shah.maih@newcastle.edu.au

ABSTRACT -

This research's primary task is to inspect the affiliation between the implementation of technology and e-commerce's success. It is imperative to study such an important relationship that directly impacts the rapid growth of Internet technology, new dimensions of e-services, and innovative measures that are necessary factors for electronic commerce (e-commerce) operations. Despite most Saudi Arabia retailers being aware of technological advancements, existing research reveals several challenges that hinder the adoption of e-commerce strategies, including the cost of installation and training. The advantages of e-commerce are frequently shown in recent studies. Internet technologies development has narrowed the difference between traditional trade and online business grounds, with additional traditional markets moving to online platforms. The Saudi Arabia community has been recognized as a potential hub for advancing technology-based programs, particularly e-commerce, given a strong GDP per capita income and a 3.4 percent annual population growth.

## I INTRODUCTION

Business technologies are the essential elements of everyday businesses and livings; computers, smartphones, and the Internet are many people's integral life parts. Effective electronic commerce (e-commerce) operation is crucial for small and medium enterprises (SMEs) in any market. The number of people who prefer to make purchases on the Internet is growing annually, where companies without e-commerce risk losing customers. Online purchases in Saudi Arabia grew from 18% to 40% between 2015 and 2017, significant [1] The growth of SMEs in developing countries is limited by poor e-commerce development compared to the developed countries using technology for a long.

This study aims to understand the relationships between adoption of technologies and e-commence success, to stimulate the e-commerce development in Riyadh, by analyzing the current situation of e-commerce both in the country and in the city in particular. One of the outcomes would be the benefits of introducing e-commerce to modern SMEs, considering the characteristics of consumers in Riyadh and Saudi Arabia by utilizing a quantitative and qualitative mixed-method approach. Previous research in this area, such as "SMEs engagement with e-commerce, e-business, and e-marketing" [2] and "the impact of effective factors on the success of e-commerce in small-and-medium-sized companies" [3], has been conducted for promoting potential benefits of e-commerce technologies in today's highly competitive markets. However, none of these addressed the situation surrounding Saudi's ecommerce development SMEs context. Due to different cultural and legal features, enterprises in Riyadh cannot simply use the experience of businesses from other countries because these approaches are not always suitable for the local environment. This study aims to fill the gaps in knowledge about the prospects and ways of developing e-commerce in Riyadh whole of Saudi Arabia. This study will contribute to bringing insightful knowledge for more effective development and strategic improvements of local SMEs, which are important to compete in both the domestic and foreign markets and support the economic development of the region.

E-commerce is a significant technology in modern businesses, enhancing connection between business stakeholders electronically, allowing SMEs to achieve higher growth [4]. Businesses should follow e-commerce-trends, requiring the adoption of relevant Internet-enabled technologies to remain competitive. SMEs are rapidly adopting various e-business methods to improve performance [5]. E-commerce improves the value chain processes, specifically production and distribution, enhancing streamlined transactions and communications.

SMEs are undoubtedly one of the major economic growth forces in all countries [6]. Economic development in Saudi Arabia benefits from increased profitability and fair competition of SMEs across the region. SMEs should, therefore, consider adopting technologies that will result in favorable business outcomes. This research proposal will focus specifically on assessing the impact of technology and it's usage ease and compatibility for e-commerce success in Saudi Arabia SMEs.

This research aims to define how technologies affect e-commerce in SMEs and what factors contribute to the successful adoption of e-commerce. The definition of success is an important factor. E-commerce success is a three-key outcome: lower customer acquisition cost, higher repurchase rate, and healthy margins. However, the outcomes are critical for Saudi Arabia's SMEs, which have grappled with digital retailing problems for the past decade. The research question is designed on the central aspects. The study attempts to specifically point out the impact of e-commerce adoption on various SMEs, including employee satisfaction, effects on enterprise operations' effectiveness, the role of e-business in decision-making processes, and the examination of existing adoption strategies for SMEs. E-commerce penetration in Saudi Arabia was only 1.4% in 2017; less than half of the US in 2005. Saudi Arabia's e-commerce is significantly behind many developed countries [7] mainly because of the laws and regulations regarding e-commerce, which should protect consumers and limit e-commerce capabilities [7]

Moreover, people are skeptical when performing online shopping because laws and regulations have significantly developed the e-commerce industry for a long. Consumers are concerned, and most of the cases, confused about the online payment process and delivery dilemmas. It is also important to note that Saudi Arabia is experiencing lower credit-card penetration and debit-cards often encounter regulatory hurdles [7]. The research is important, considering SMEs make up about 90% of Saudi Arabia's registered businesses [8].

Specifically, the research will address the problem of the slow adoption of online retailing SMEs in terms of business to customers (B2C) in Riyadh, Saudi Arabia, by posing the following research questions:

- What are the influences that force Riyadh businesses towards the implementation e-commerce technologies?
- What influences the effective implementation of e-commerce strategies' technologies?

This study aims at filling a gap caused by inadequate information in this area and enhance the study that links technology and e-commerce. The study conveys new prospects to Riyadh SMEs by determining the affiliation between technology embracing and effects on business' results. Therefore, Riyadh SMEs managers can obtain a new understanding of how e-commerce technologies are important for their business practices. Moreover, new attributes will be studied and added to existing technology adoption theories, making recommendations to Riyadh SMEs. The recommendations enhance e-commerce technologies' introduction into organizations, critical in today's highly competitive markets.

Additionally, this research undertakes the Diffusion of Innovation (DOI) conceptual model, developed by [9] explaining how, why, and at what speed new ideas and technologies are spreading. According to the theory, innovation must be widely accepted to support itself. Various fundamentals inspire the growth of new ideas. Innovation depends on communication networks, period, and social system reaction [9]. Besides, the innovative process is dependent on human capital. Therefore, innovation must be widely accepted by human users to support itself. Also, within the limits of the adoption norm, there is always a moment when innovation reaches its critical mass, where further knowledge dissemination is impossible (Rogers, 1962).

Finally, this study's purpose is the general strengthening of knowledge within the discipline of information systems, and in particular, an understanding of the process of introducing e-commerce and all the features of its implementation in the modern business and political environment of Saudi Arabia.

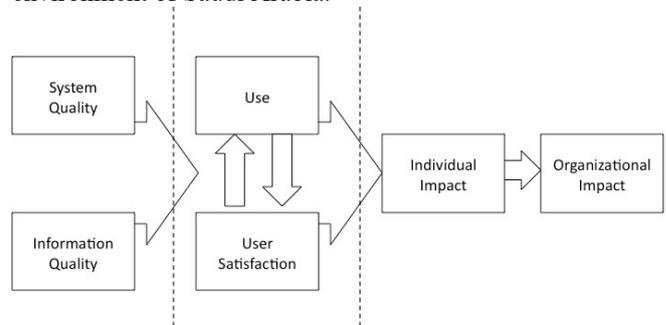

*Figure 1: Information systems success model [10] which is the central theory of this study*

More specifically, according to the Information systems success model (ISS) by [10], there are six most critical aspects of success for information systems, namely; information, system and services qualities, system usage intents, user gratification and overall system returns [10]. The aspects are interconnected with each other; hence, the flaws can affect the entire system. For the system to be successful, when introducing e-commerce, it is imperative to consider each factor. This study therefore examines these aspects of success in terms of e-commerce and complement existing knowledge in this area.

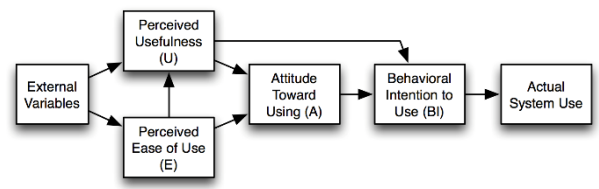

*Figure 2: TAM model [11] which is used as a sub-model for understanding technology acceptance*

Besides, this study uses the technology acceptance model (TAM), which models the process as users accept and use it. According to this model, several factors directly or indirectly influence users' decisions about technology usage [11]. The model's author identified two key factors, perceived utility and ease of use. Therefore, a particular system's users must improve their productivity and be easy to use [11]. The study will combine ISS and TAM to understand the main factors of e-commerce implementation, making it possible to give recommendations to enterprises.

This quantitative study considers two independent variables (firm size and adoption of technology). It will have ease of use of technology as the moderating variable and compatibility as the mediating variable. The dependent variable of this study will be the success of the e-commerce strategy. In terms of period, this study will be conducted within three months for collecting field data. The target population for this study will be managers of SMEs in Riyadh. The stratified sampling method will be used, and data will be collected through semi-structured questionnaires addressed to ICT experts in the enterprises. Such interviews contain specific thematic requiring coverage, but the questions' order wording is left to the interviewer's direction [12]. This approach's key feature is that the researcher can answer questions from participants, ask questions, and discuss issues of importance to them.

Additionally, the study aims to add more information regarding social media integration in Saudi Arabian businesses and how to widen their customer base.

Moreover, the study's results should understand Saudi SMEs about what steps should be taken to gradually and efficiently introduce e-commerce technologies in existing and new enterprises and avoid potential problems. It is vital to recognize that despite the willingness of Saudi Arabians to develop e-commerce in the country, SMEs must find a unique approach for their target audience so that potential customers have a positive online consumer experience and share it with others. Further investments in the region and economic stability are also critical.

This study aims at finding out the benefits of adopting e-commerce technology in Saudi Arabian SMEs. The importance of technology in improving performance in modern businesses cannot be ignored. This study will fill a knowledge gap concerning e-commerce technologies in SMEs. While carrying out the study, technology will be used as the moderating variable, while its compatibility with the SMEs will be the mediating variable.

## II LITERATURE REVIEW

The e-commerce domain has been rapidly developing over the past few decades. The literature in this domain has provided the latest understanding and knowledge for operating effective online businesses. Understanding the rewards of e-commerce and suitable methods for realizing success are significant for B2C businesses across the globe.

Many studies discourse ecommerce effects and benefits in business. For instance, [13] compares diverse e-commerce definitions, giving a wide-ranging assessment of its meaning in business. Another article [14] highlights important e-commerce development trends in Saudi Arabia, offering insight into its state and comparing it with similar technologies. On the other hand, [15] looks into technology in e-commerce. [15] looks at the impact of technology on e-commerce. Studies look into the influence of IT tools on e-marketing. It is also important to mention [5] and [6], address factors that influence SMEs' e-commerce adoption. Lastly, the literature review addresses the disadvantages and advantages of e-commerce acceptance.

E-commerce has attained more attention because of the its transformative effects upon businesses. Many researchers have looked into the importance of e-commerce technology to business organizations. However, there exist many restrictions that deter our understanding of the whole picture. This study's objective is to produce a detailed review of current studies and undertake quantitative exploration that will contribute to the knowledge about e-commerce technology.

*A. 2.2. E-commerce in the current business environment*

There have been numerous attentions on e-commerce in business, enabling the promotion of products hence better results. E-commerce has enhanced competition between online businesses. The difference between traditional and online retail has been reduced, as more traditional retails shops have switched to the use of online platforms [13]. Electronic devices use has also enhanced the use of e-commerce technologies.

The objective of e-commerce is introduction of a better environment where businesses can sell and market their products. [13] defines e-commerce as the different online tools and business activities used by companies to achieve their strategic objectives of selling products and services. However, the definition, does not define the specifics of e-commerce sufficiently. Understanding e-commerce needs a more detailed explanation of its primary purpose and tools. E-business is used to enhance a business per se despite the provisions of the necessary tools for organization to organization relationship improvement by e-commerce. Table 1 illustrates a critical analysis of how the gaps are outlined for the proposed study [13]

*Table 1: Critical analysis of the previous studies of e-commerce*

| Studies | The current literature review assessed | Articulated research gaps for the current research |
|---|---|---|
| [14] | significant trends in e-commerce development insight into the state of it in the country and comparing it to other retailing technologies. | The data presented will be used as the basis for determining key performance indicators (KPIs) in assessing the success of e-commerce implementation. Further analysis also provides recommendations to SMEs in Riyadh when adopting technologies. |
| [16] | Sixty-five percent of Saudi Arabia's population has access to the Internet; online shoppers in the country have rapidly increased. | We noted that one of the important factors contributing to the development of e-commerce is the ease of use. The relationship between ease of use and the success of e-commerce will be explored. |
| [16] | User readiness and awareness lack of trust, online modes of payment, internet privacy safety and credit card security are key inhibitors to the gradual adoption of online retailing. | This is a general description of possible barriers towards e-commerce implementation, tough there exist more profound description about businesses challenges. A more in-depth analysis of potential issues in adopting e-commerce will help identify potential success factors, which is |

| | | an important part of the research. |
|---|---|---|
| [17] | It examines the features of advertising needed to promote e-commerce after implementing information technology. | Advertising is undoubtedly one of the tools used to increase the effectiveness of any process on the Internet. |
| [18] | It studies the relationship between social networks and the development of e-commerce as integral elements in modern markets. They offer a new solution for the cross-site cold-start recommendation. | The authors determined that social networks are also important factors affecting the speed of technology distribution and its relevance. |
| [19] | They examined how culture influences trust in e-commerce. National culture can directly influence perceived trust. | Ease of use will be used as one of the factors affecting consumers' desire to use the service. It is necessary to identify and understand potential consumers' needs |

Various statistical reports and studies show an increased integration of mobile and Internet technology in Riyadh, Saudi Arabia, over the last few years. This section illustrates studies that depict Saudi Arabia's demographics and resources and how they influence e-commerce adoption in the country. Riyadh's e-commerce market's growth is mainly attributed to the nation's GDP figures at one hundred and sixty-eight billion dollars making Saudi Arabia the largest economy throughout the Arab region [20]. Through its strong GDP per capita income and a 3.4 percent annual population growth (Al-Mosa, 2011), the Saudi Arabia community is a potential hub for advancing technology-based programs, particularly e-commerce.

E-commerce in Saudi Arabia has gained significant attention leading to numerous studies and research on the topic. [21] surveyed online shoppers to understand the aspects of e-commerce acceptance in Riyadh. Furthermore, [16] reports that sixty-five percent of Saudi Arabia's population has access to the Internet. However, the report does not include details of the transactions made by related parties. According to a research report by [16], various factors such as user readiness and awareness, lack of trust, online modes of payment, internet privacy and safety, and credit card security are key inhibitors to the gradual adoption of online retailing.

The government and private sectors are making efforts towards the improvement of e-commerce adoption. The Saudi government has started initiative towards the encouragement of citizens on suing online platforms. The initiatives enhance the integration of offline services like visa, traffic and public utility services to online platforms. The therefore can citizens benefit from faster and more convenient public services. Consumers and retailers are however affected by the massive control of internet and related services by the government [22] hinedring e-commerce development. However, enhancing individual and business security and implementation of e-commerce – specifically encryption measures –can meaningfully promote e-commerce infrastructure in the Saudi.

An efficient e-commerce infrastructure requires the combination of the necessary tools and technologies. The aspect ensures that the infrastructure have all the abilities to enhance the customer activities on a platform hence improving their satisfaction. One of the most implemented e-commerce technologies is mobile apps which are enormously suitable for consumers and retailers. The apps ensure that consumers do not have to visit stores physically which saves their time. Additionally, the applications bring more benefits to retailers and consumers.

Business newsletters are other tools which enable communication between the retailer and consumers about new products and services. Companies are adopting the method for offering personalized client experience. The activities improve the customer loyalty [23]. Organizations also have to follow the emerging trends to enhance their competitiveness. Some of the most used technologies are social media platforms such as Instagram, Facebook, and Twitter. The platforms are powerful for product promotion and marketing due to the vast amount of time consumed by people on the platforms. The trends can enhance the development of Saudi SMEs

There is a gap in this knowledge that will be covered in this research in the existing studies. There is no study that adequately looks into the e-commerce developments (particularly for SMEs) in Saudi Arabia. There is a need to conduct more in-depth research to address the areas of e-commerce development in Riyadh and trends present among Saudi Arabia's SMEs. This will encourage more business to integrate modern technologies, that will aid them achieve greater business results by addressing dynamic competitive global-markets and contributing to the country's economic growth. [24] noted that government regulation and support were the major challenges in introducing e-commerce in Saudi Arabia. In turn, our study will focus solely on the technology adoption of SMEs to fill the knowledge gap.

The importance of technology and various insights in e-commerce are addressed. Various definitions are focused on. Moreover, the primary trends in technology and e-commerce in Saudi Arabia are described to describe e-retailing in the country. The review addresses the topics of the effects of e-commerce strategy on the success of the business. Lastly, the gap in the present literature is identified to address this gap in this research.

## III METHODOLOGY AND CONCEPTUAL FRAMEWORK

Defining the study's methodology is an essential step in researching because the chosen method can largely influence the study's primary findings. Asit determines how a researcher gathers and interprets data. In this chapter, the central concepts of this research methodology are described and explained. This research's accepted paradigm is positivism, which focuses on scientific and accurate data for deriving conclusions. It is then essential to choose the research methodology, whether it is qualitative, quantitative, or mixed. In this research, a mixed methodology will be used to predominance a quantitative method for data analysis. In the first part of the study, it will be necessary to use a qualitative method for the development of the questionnaire, which will then be sent to small and medium-sized enterprises of Riyadh, which will be a source of data for further analysis. Lastly, the chapter provides an insight into the primary ethical considerations of the study.

This research will adopt a critical paradigm. Critical paradigm focuses on the experience of other people and their perceptions, so this is the most accurate method for finding the relationship between the adoption of technology and eCommerce's success. This requires studying other people's experience in this field because even implementing the same project in different countries and industries can have completely different results. The questionnaire, which will be developed for enterprises in Riyadh, will provide the researcher with subjective information on introducing e-commerce. This information will be grouped and analyzed to understand the similarities and differences between the respondents' answers. In this study, subjectivity is essential to provide readers and entrepreneurs with an adaptation of the effectiveness of introducing e-commerce by other enterprises that have already achieved some success, which is also useful for further studying the topic.

Although some elements of several paradigms can be used in this research, the primary emphasis is on the critical paradigm, which focuses on other people's experiences and perceptions. This research uses methods for developing and testing the hypotheses that will significantly impact the results of the study. The four hypotheses that will be tested in this research are the following:

**1. There is a significant relationship between the technology adoption stage and success in e-commerce.**
The technology adoption life cycle is a sociological model that describes the adoption of a new product or innovation following certain user groups' demographic and psychological characteristics [28]. According to this theory, the first group of people using the new product is called "innovators," followed by "early adopters." This is followed by an early majority and a later majority, and the last group that will ultimately take the product is called Laggards or Phobics [28]. This hypothesis suggests that if the technology at this moment is in the early stages when it is in demand only among innovators and early adopters, it has less chance of success [29]

**2. There is a significant relationship between ease of use and success in e-commerce.**
Undoubtedly, clients are one of the key values of companies; therefore, it is assumed that their satisfaction positively affects the success of a project, product, or organization as a whole. According to [30] and the technology acceptance model, ease of use is one of the key factors affecting customer satisfaction. According to [3], which conducted a study to identify key factors that influence e-commerce success, have confirmed the close relationship between customer satisfaction and e-commerce success.

**3. Technology compatibility contributes to the success of e-commerce in the organization.**
Technology compatibility, in turn, is necessary for a system to function successfully and efficiently [31]. Technology compatibility implies that for the successful implementation of an electronic system, it is required to have a developed infrastructure, appropriate hardware and software, and a stable Internet connection [31]. If, for example, the hardware is not fully compatible with the software, then the system will work with malfunctions and errors, which, in turn, will adversely affect customer satisfaction. [3], in their study, noted that strong and sustainable e-commerce requires an appropriate technical infrastructure.

**4. The larger the size of a company, the bigger the impact of e-commerce on business performance.**
The introduction and support of e-commerce in the organization undoubtedly require certain costs and long-term investments. The costs of introducing e-commerce are practically independent of the company's size and the number of customers because they are constant and only increase slightly with increasing company size [32]. Since e-commerce in Saudi Arabia is still underdeveloped, it is impossible to say exactly how much it will be in demand among certain consumers. Accordingly, companies cannot know whether their investment will pay off; in the case of a large company, the shutdown can significantly impact the activities of the company, while inefficient investments in a small company can stop the development of the business. [3] noted that high costs weaken the success of e-commerce in an organization. The proposed hypothesis suggests that the larger the company, the greater the likelihood of successfully implementing e-commerce technology.

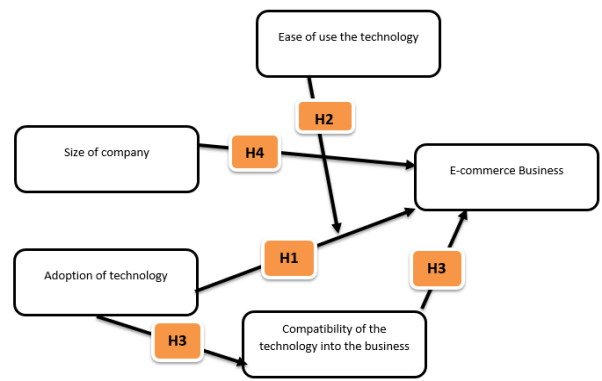

*Figure 3: A proposed conceptual model*

In figure 1, the moderator of the variables is the ease of use of technology, technology eases of use is the variables' moderator because it affects the link between technology adoption and e-commerce. The compatibility of technology in a business, which explains why the two variables exist is the mediator.

*Table 2: Measurement table for the variables*

| Variable | Measurement |
|---|---|
| Ease of use | 1. Completion Rates<br>2. Usability Problems<br>3. Task Level Satisfaction<br>4. Expectation<br>5. Page Views/Clicks<br>6. Conversion. |
| Compatibility of technology in a business | Compatibility of business processes and company needs with e-commerce technology |
| Size of a company | Small, medium |

| Adoption of technology | Innovators, Early Adopters, Early Majority, Late Majority, Holdouts (Rogers, 1962). |
|---|---|

There is also a mixed methodology that combines both qualitative and quantitative approaches to conducting research. Although the qualitative methodology is widely used in social research, quantitative research can also be applied to behavioral and social studies, bringing valuable results. Furthermore, both methodologies can be highly useful for making an in-depth analysis of the subject and considering both sides of the research.

Riyadh Chamber of Commerce has a list of all small and medium enterprises containing 2100 records. Our process involves contacting the businesses by email and ask them to fill out a questionnaire, which will explain the current situation of e-commerce in Riyadh. It is expected that 120 companies will fill out a questionnaire, and this data will be used for further analysis. After the companies fill out the questionnaires, we will have information about which companies use e-commerce and then compare them with their profitability. To obtain reliable data that can be used to further study the issue, interviews, and at least ten different enterprises that operate in various industries and have different experiences and outcomes will be conducted. To avoid bias, it is necessary to conduct interviews with companies that can share their experiences from different perspectives. This will ensure that the study results are reliable and useful both for further researchers on this topic and for enterprises in Riyadh.

The study used primary quantitative data, which will be obtained through interviews and structured questionnaires on the impact of technology adoption on the effectiveness of e-commerce strategies in Riyadh SMEs.
The first stage of this study will be qualitative interviews with ten SMEs enterprises in Riyadh. This study will collect data through semi-structured in-depth interviews with top managers of enterprises. Such interviews have specific thematic areas that will need to be covered during the interview. Still, the order of questions and their wording is left to the interviewer's discretion [12]. A key feature of this approach is the fact that it allows the researcher to answer questions that may arise from participants, ask questions, and also allows participants to discuss issues that are considered important to them.

For interviews with representatives of small and medium enterprises in Saudi Arabia to be as informative and useful as possible for further analysis, it is needed to develop a list of questions in advance. This will mainly be based on a study called 'SMEs' adoption of e-commerce using social media in a Saudi Arabian context: A systematic literature review. " In this study, the authors accurately described all the gaps in knowledge that need to be filled to understand at what stage of development e-commerce is currently in Riyadh in particular and Saudi Arabia as a whole. After the interview questions have been developed, it will be necessary to send these questions by email to the interviewees to develop tentative answers, which will greatly facilitate communication.

During the interview, the researcher will need to learn about all the features of the development of electronic commerce in Saudi Arabia and study how these enterprises were able to implement this information technology, what difficulties they encountered and what results they got.

The questionnaire will comprise structured questions drawn from previous empirical studies, interviews, theory, and the researcher's questions based on the context. The questionnaire will mainly be based on the results of the studies, "An investigation of the impact of effective factors on the success of e-commerce in small- and medium-sized companies" and "SMEs' adoption of e-commerce using social media in a Saudi Arabian context: A systematic literature review. ". The authors studied the features of introducing e-commerce in small and medium enterprises [3]. The questions will be designed to supplement the missing knowledge and understand the features of e-commerce in Riyadh. Moreover, the questions will be compiled based on interviews conducted with representatives of small and medium-sized corporations in Riyadh.

the researcher will ask experts or people who understand the topic under study to read my questionnaire. They will need to evaluate how effectively questions cover the topic. Moreover, the researcher will ask the psychometrics to check the survey for common errors, such as confusing, suggestive, and two-dimensional questions. After that, it will be necessary to test the questionnaire among a small number of companies to assess which issues are necessary and effective. On the contrary, they are useless. After collecting the pilot answers, the researcher will enter the spreadsheet's answers and check how rashly the participants filled out the questionnaire.

IV RESEARCH DEVELOPMENT

The study has created questions that must be researched and answered. First of all, five of the participants disagreed that sales have improved due to e-commerce implementation, which should not be the case if it is implemented correctly. More data must be gathered to understand why e-commerce works for certain organizations and why it does not work for others—another increased customer satisfaction. E-commerce has proven to be effective for the world's biggest companies, so each instance of dissatisfaction must be thoroughly researched with a narrow scope. Other latest technologies [35,36,37] and their potential impacts on the success of SMEs also can be another relevant study for future, following design theories [38].

V CONCLUSION

Due to the rapid development and spread of the Internet, people are increasingly using it in everyday tasks, and businesses are finding ways to monetize it. The Internet has contributed to the explosive success of e-commerce in the West, which has led to the world's wealthiest person owning an online store. Due to legal and certain consumer restrictions and lag, e-commerce in Saudi Arabia has not received such development as in the USA or Europe. But in recent years, laws have been changed, and credit cards with other flexible methods of payments have become more popular among residents; consumer interest and demand have grown continuously, which, together with many other factors, has contributed to the spread in Saudi Arabia. There is a steady trend for Internet sales, and companies in Saudi Arabia, in particular Riyadh, should use this opportunity to develop their business.